\documentstyle[aps,eqsecnum,amssymb,twocolumn,floats,epsf]{revtex}

\begin{document}
\draft

\title{Mean field approach to antiferromagnetic domains
in the doped Hubbard model}
\author{Edwin Langmann and Mats Wallin}
\address{
Department of Theoretical Physics, Royal Institute of Technology,
S-100 44 Stockholm, Sweden \\
\date{\today}
}

\newcommand{\intdk}{\int\! d\!\!^{\mbox{-}}\!\,{\bf k}\, }
\newcommand{\Ln}{{\rm Ln}_\beta}
\newcommand{\vk}{{\bf k}}
\newcommand{\vs}{{\bf s}}
\newcommand{\vQ}{{\bf Q}}
\newcommand{\vx}{{\bf x}}
\newcommand{\vy}{{\bf y}}
\newcommand{\vm}{{\bf m}}
\newcommand{\vn}{{\bf n}}
\newcommand{\ve}{{\bf e}}
\newcommand{\vsigma}{\mbox{\boldmath $\sigma$}}
\newcommand{\vphi}{\mbox{\boldmath $\phi$}}
\newcommand{\eq}{\begin{equation}}
\newcommand{\eqend}{\end{equation}}
\newcommand{\eqa}{\begin{eqnarray}}
\newcommand{\nonueqa}{\begin{eqnarray*}}
\newcommand{\eqaend}{\end{eqnarray}}
\newcommand{\nonueqaend}{\end{eqnarray*}}
\newcommand{\nonu}{\nonumber \\ \nopagebreak}
\newcommand{\bma}[1]{\begin{array}{#1}}
\newcommand{\ema}{\end{array}}
\newcommand{\bc}{\begin{center}}
\newcommand{\ec}{\end{center}}
\newcommand{\R}{{\Bbb R}}
\newcommand{\Z}{{\Bbb Z}}
\newcommand{\cF}{{\cal F}}
\newcommand{\xx}{{\rm x}}
\newcommand{\Tst}{T^\star}

\maketitle

\begin{abstract}
We present a restricted path integral approach to the 2D and 3D repulsive
Hubbard model.  In this approach the partition function is approximated by
restricting the summation over all states to a (small) subclass which is
chosen such as to well represent the important states.  This procedure
generalizes mean field theory and can be systematically improved by
including more states or fluctuations.  We analyze in detail the simplest
of these approximations which corresponds to summing over states with local
antiferromagnetic (AF) order.  If in the states considered the AF order
changes sufficiently little in space and time, the path integral becomes a
finite dimensional integral for which the saddle point evaluation is exact.
This leads to generalized mean field equations allowing for the possibility
of more than one relevant saddle points.  In a big parameter regime (both
in temperature and filling), we find that this integral has {\em
two} relevant saddle points, one corresponding to finite AF order and the
other without.  These degenerate saddle points describe a phase of AF
ordered fermions coexisting with free, metallic fermions.  We argue that
this mixed phase is a simple mean field description of a variety of
possible inhomogeneous states, appropriate on length scales where these
states appear homogeneous.  We sketch systematic refinements of this
approximation which can give more detailed descriptions of the system.
\end{abstract}


\section{Introduction}

The two dimensional (2D) 1-band Hubbard model \cite{Hubbard} has
received much attention as a simple prototype model for high temperature
superconductors (HTSC) \cite{Anderson}.  It provides a good
description of the insulating antiferromagnets which, upon doping, become
HTSC.  Of central interest is the transition from the insulating
antiferromagnetic (AF) state to the metal state at high doping.
Experiments on HTSC show that AF correlations are also important away from
half filling, but despite of much effort a satisfactory theoretical
understanding of HTSC based on the Hubbard model has not been achieved (for
a recent review see \cite{Review}).

Hartree-Fock (HF) theory is the basis of most successful theories in solid
state physics: `\ldots {\em modern many-body theory has mostly just served
to show us how, where, and when to use HF theory and how flexible and
useful a technique it can be}' \cite{Anderson0}.  One should thus expect
that it is useful also for correlated electron systems.  Indeed, HF theory
seems is a good starting point for describing the half filled Hubbard
model.  For strong coupling, it predicts a ground state with AF long range
order (= N\'eel state) \cite{BL}.  This state is invariant by translations
by two sites and thus allows a simple analytic treatment using Fourier
transformation.  Away from half filling, HF theory of the Hubbard
model becomes very complicated.  The reason for this is an apparent
rigidity of the AF ordered electron system which does not allow
homogeneous doping.  A variety of
inhomogeneous HF ground states have been found, including magnetic domain
walls \cite{walls}, magnetic polarons, magnetic vortices
\cite{HartreeFock}, spiral states \cite{spiral}, and phase separation
\cite{phasesep}.  Systematic investigations of these states have been done
by numeric methods and were restricted to small dopings and finite lattices
\cite{VLLGB}. In these calculations the HF equations
for general space dependent HF fields are analyzed.  Denoting
(imaginary) time as $\tau$ and lattice points as ${\bf x}$, these HF
fields are
\begin{equation}
\label{Neel1}
\phi_0(x)=r(x),\quad \mbox{\boldmath $\phi$}(x) = s(x){\bf e}(x)\,
e^{i\vQ\cdot\vx}
\end{equation}
[we write $x=(\tau,{\bf x})$, and our lattice points are ${\bf
x}=(x_1,\ldots,x_d)$ with $x_i$ integers] where $s$ is the magnitude of the
{\em local} AF order parameter, ${\bf e}$ its direction, and $r$
is a charge degree of freedom; $\vQ=(\pi,\ldots,\pi)$ is the
AF vector as usual.  In the doped region, one finds a large number of
solutions of the HF equations which correspond to metastable spin and
charge configurations \cite{VLLGB}.

Such numeric calculations are important to learn what the essential
features of the Hubbard model are.  Ultimately, however, it would be
desirable to develop analytic methods taking into account these essential
features and allowing simple computations.  In this paper we propose such a
method which corresponds to an averaging over highly degenerate HF
solutions.  It leads to a simple homogeneous mean field description of AF
in the doped Hubbard model which makes manifest the physical expectation
that, at large enough length scales, the system should appear homogeneous
even if the HF solutions are not.  The results of this paper are restricted
to the simplest non-trivial ansatz for states describing AF order.  We find
that already this approximation allows to systematically determine the
parameter regime where no homogeneous AF solutions of the HF equations are
possible, and it leads to a simple understanding of this phenomenon.  Our
approach can be interpreted as approximation of the exact partition
function by restricting the sum over the (huge) set of all states to a
(small) subset which one chooses such as to to well represent the `most
important states'.  This approach is very general and can be systematically
refined by increasing the subset of states taken into account.  Thus one
can obtain increasingly complicated approximations which will give
increasingly refined information about the dominating states of the system.
Moreover, also fluctuation can be naturally taken into account in this
approach (this will be only outlined in this paper).

It is often believed that mean field theory (based on HF equations) is not
appropriate for correlated fermions systems.  We propose that it is only
mean field theory {\em in its usual formulation} which fails.  Our approach
leads to a generalized mean field theory which, as we believe, is a
reasonable starting point for a good theory of these systems also.
Fluctuations are important, of course (especially in two dimensions), but
are corrections to generalized mean field theory, as discussed.

We now describe our method in more detail.  We recall that the HF fields
$\phi(x)=\bigl(\phi_0(x),\mbox{\boldmath $\phi$}(x)\bigr)$ in Eq.\
(\ref{Neel1}) naturally appear if one writes the exact partition function of
the Hubbard model as a boson path integral using a Hubbard-Stratonovitch
transformation, $Z = \int{\cal D}\phi \,
\exp[-\cF(\phi)]$ \cite{Schulz}. The boson action $\cF(\phi)$ here
has a simple physical interpretation: it is the free energy of
non-interacting fermions in the {\em external} boson fields $\phi(x)$.  HF
theory amounts to a saddle point evaluation of this path integral
\cite{Wolff}.  Since $\cF(\phi)$ is a complicated functional of $\phi$, a
search of the relevant saddle points among all possible spacetime dependent
boson field configurations $\phi(x)$ is not feasible.  Thus usually a
certain ansatz for $\phi(x)$ is made.  We will refer to this procedure as
{\em mean field theory}: the search for saddle points of $\cF(\phi)$
restricted to a (small) subset of boson configurations.  Obviously one can
expect this to give a good description of the system only of this subset is
chosen such as to contain boson configurations which are sufficiently
similar to the saddle points which actually dominate the exact path
integral.  We now propose: Instead of analyzing the saddle point equations
by restricting to some subset of boson fields, {\em approximate the HS path
integral by summing over all boson configurations in this subset.}
Parameterizing this subset by finitely many real parameters, we thus
approximate $Z$ by a finite dimensional integral.  In many interesting
cases one finds that a saddle point evaluation of this integral is exact,
and we recover standard mean field theory {\em in case there is only one
relevant saddle point}.  In general, we obtain a generalized mean field
theory allowing for degenerate saddle points to contribute to the partition
function.  To be more specific, for the Hubbard model the simplest ansatz
for boson fields Eq.\ (\ref{Neel1}) describing AF order is
\begin{equation}
\label{Neel}
\phi_0(x)=r,\quad \mbox{\boldmath $\phi$}(x) = s\, {\bf
e}\,e^{{i}\vQ\cdot\vx }\: .
\end{equation}
Eq.\ (\ref{Neel}) describes a N\'eel state i.e.\ AF long range order.  This
ansatz gives a reasonable description of the Hubbard model at half filling,
thus it is natural to also attempt it in the doped regime.  One can sum
over all configurations of the form Eq.\ (\ref{Neel}) and thus approximate
$Z$ by an integral over $r$, $s$ and ${\bf e}$.  Since $\cF$ in this case
is proportional to the spacetime volume, $\cF=\beta L^d f_0(r,s)$ [$L^d$ is
the number of lattice sites, $f_0$ a function given in Eq.\
(\ref{f0}) below], and independent of ${\bf e}$, and the free energy per
lattice
site is $\Omega= -\log(Z)/\beta L^d$, we obtain
\begin{equation}
\label{Om}
e^{-\beta L^d \Omega }  = \int_0 ^\infty d  s\,
\int_{-\infty}^\infty d  r\, e^{- \beta L^d f_0(r,s)}
\end{equation}
(the ${\bf e}$-integration gives an irrelevant constant factor which we
drop).  In the thermodynamic limit $L\to\infty$, the saddle point
evaluation of this integral is exact.

We now discuss in more depth the nature of this approximation which is
necessary to correctly interpret our results described below (this
paragraph is not essential for understanding these results, however).  It
is important to note that our Eq.\ (\ref{Om}) is not only obtained by summing
over the N\'eel states Eq.\ (\ref{Neel}), but in fact also by summing over
a much larger class of configurations Eq.\ (\ref{Neel1}) where $r(x)$, $s(x)$
and ${\bf e}(x)$ are `slowly' varying functions with `fast' changes
occurring only in a small fraction of the total spacetime volume: there are
such configurations {\em without long-range order} but still
such that they {\em locally} resemble a N\'eel configurations Eq.\
(\ref{Neel}) so much that adding their contribution to the partition
functions essentially leaves the result unchanged.  These configurations
describe states with {\em local} AF order (i.e.\ they have a finite
correlation length).  For these configurations, the boson action can be
computed as $\cF=\cF_0+\cF_1$ with
\eq
\label{HSaction}
\cF_0(\phi) = \sum_x f_0\bigl(r(x),s(x)\bigr)
\eqend
and $\cF_1$ a correction involving gradient terms which are negligibly
small if $r(x)$, $s(x)$ and ${\bf e}(x)$ change sufficiently little.  This
formula has simple physical interpretation: for states with local AF order,
each spacetime point $x$ contributes a term $f_0\bigl(r(x),s(x)\bigr)$ to
the fermion free energy.  This term depends only on the local $s$- and
$r$-values (the local direction of the AF order is irrelevant), and thus is
the same one would have in a N\'eel state, provided $r(x)$, $s(x)$ and
${\bf e}(x)$ have sufficiently little variation.  This argument immediately
shows that Eq.\ (\ref{Om}) is also obtained if we sum over all boson
configurations Eq.\ (\ref{Neel1}) with $s(x)=s$ and $r(x)=r$ constant but
${\bf e}(x)$ varying not much as a function of $x$.  More generally, we
will show that also $s(x)$ and $r(x)$ are allowed to vary, provided the
average correlation volume where $s(x)$ and $r(x)$ are constant is
sufficiently large: if we sum over such configurations, we also obtain
equation (\ref{Om}) but with $\beta L^d$ replaced by this correlation volume.
To avoid misunderstanding, we stress that this does not mean that the
effect of fluctuations, domain walls etc.\ is not important: this argument
only shows that a nontrivial solution of HF equations (i.e.\ with $s\neq
0$) resulting from Eq.\ (\ref{Om}) should not be interpreted as N\'eel order
but as existence of {\em local} AF correlations.  To decide weather there
is long range order or not one has to go beyond this simple approximation.
For example, one can improve Eq.\ (\ref{Om}) by making a more general ansatz
for the boson configurations $\phi(x)$ to be summed over (describing e.g.\
domain walls etc.), or one can include the effect of fluctuations.  Only
this will give more detailed information about the structure of the state.

A main result of this paper is that, in a big parameter regime, the 
integral in Eq.\ (\ref{Om}) has two relevant saddle points: one with 
$s=s^*>0$ corresponding to (local) AF order, and another with $s=0$ 
corresponding to no AF order.  This means that no HF solution of the form 
Eq.\ (\ref{Neel}) exists and, as discussed below, unusual physical behavior 
is to be expected.  Physically this can be interpreted as follows: the 
homogeneous AF ordered electron system can only exist at half fillings, and 
the only way to achieve doping is to have the chemical potential such that 
the AF saddle point is degenerate with a trivial one which serves as a 
particle reservoir.  The analysis of the integral (\ref{Om}) provides a 
simple and systematic way to determine the parameter regime where this 
happens.  The resulting phase diagrams for the Hubbard model and different 
parameter values is shows in Figs.~3 and 4.  There are always three 
different regimes: a AF regime where the non-trivial saddle points 
$s=s^*>0$ dominates, a free regime with the only relevant saddle point 
$s=0$, and the mixed regime discussed above.  The doping $\xx_1$ separating 
the AF regime from the mixed one increases with temperature it approaches 
the doping $\xx_2$ separating the mixed and the free regime.  At a 
characteristic temperature $T^{\star}$, $\xx_1$ merges with $\xx_2$ and the 
mixed regime disappears.  Note that since $\xx_2>0$, the free regime 
describes metallic fermions, i.e.\ there is a Fermi surface with a finite 
density of states (DOS).  For low temperatures, $\xx_1\approx 0$ i.e.\ the 
AF regime is insulating (no DOS at the Fermi surface).

It it known that, in various limits, holes in the $t-J$ and Hubbard model
accumulate in hole-rich regions separated from pure AF regions, and it has
been conjectured that this tendency to phase separation is a general
feature of holes in an antiferromagnet \cite{phasesep,EK}.  Our method can
be regarded as a simple systematic way to check this conjecture, without
any restriction of parameters: occurrence of two degenerate saddle points
shows that there is a frustration not allowing a homogeneous distribution
of holes.  However, since in Eq.\ (\ref{Om}) the effect of phase boundaries
is not included, no information about the actual distribution of the holes
is obtained that way.  Previously it has been suggested that the size of
pure-phase regions is limited only by long-range Coulomb forces \cite{EK}.
However, numeric results \cite{HartreeFock,Review} show that this happens
also without long-range forces.  Within our formalism, a more detailed
understanding of the distribution of the two different phases can be
obtained by summing over a larger class of boson configurations, e.g.\
configurations describing domain walls of variable size.

We finally note that our approach not only provides phase diagrams but also
gives approximations for all observables in the system: An exact formula
for all Green functions $G$ of the Hubbard model is given by the path
integral $G=\int{\cal D}\phi \,\exp[-\cF(\phi)] \, G_0(\phi)/Z$ where
$G_0(\phi)$ is the corresponding Green function of {\em non-interacting}
fermions in the external boson field $\phi$.  Thus by our restricted path
integral method we obtain approximations of $G$ by finite dimensional
integrals.  In case this integral is dominated only by one saddle point, we
obtain $G=G_0(\phi_{saddle})$ i.e.\ Green functions of non-interacting
fermions.  This is a simple description of a Fermi liquid.  For the mixed
phase discussed above we get
\eq
\label{1}
G=(1-w)\, G_0(\phi_{AF}) + w\, G_0(\phi_{triv})
\eqend
with a weight factor $w$ changing smoothly from $w=0$ at doping $\xx=\xx_1$
to $w=1$ at $\xx=\xx_2$.  Here $G_0(\phi_{AF})$ [$\phi_{AF}$ the N\'eel
state Eq.\ (\ref{Neel}) with $s=s^*$] describes AF ordered fermions which are
`heavy', and $G_0(\phi_{triv})$ [$\phi_{triv}$ the configuration
Eq.\ (\ref{Neel}) with $s=0$] describes free, metallic fermions.  Thus Eq.\
(\ref{1}) gives a simple description of a system where two different kinds
of fermions coexist.  This system does not resemble a conventional Fermi
liquid.  We stress, however, that Eq.\ (\ref{1}) only gives a good
quantitative description in case the effect of phase boundaries can be
neglected, and it only is appropriate on length scales where the two phases
appear homogeneously mixed.  Thus this description applies in case of phase
separation \cite{EK}.  If there is a doping regime of HTSC where this
description is adequate, the physical properties there should be of two
different coexisting kinds of fermions.  There is quite some experimental
evidence for this, as discussed in \cite{EK}.

The plan of this paper is as follows.  In the next section we describe our
path integral formalism, stressing a few important technical details.  In
Section III our approximation Eq.\ (\ref{Om}) to the partition function is
derived, and we outline how to obtain refinements to this approximation.
In Section IV we analyze Eq.\ (\ref{Om}) and calculate the resulting phase
diagram of the 2D and 3D Hubbard model for a few representative parameter
values.  Section V contains an outline of how to include fluctuations.  We
end with conclusions in Section VI.  Some technical details are deferred to
two appendices.

\section{Path integral formalism}
In this section we review the path integral formalism for the Hubbard model
\cite{Wolff,NO}, concentrating on some important technical details in
which we deviate from standard approaches.

Configuration space is labeled by $x=(\tau,\vx)$ where $0 \leq \tau\leq
\beta$ is the usual imaginary time and $\vx$ are vectors in space
$\Lambda_L$.  Our space $\Lambda_L$ is a subset of the $d$-dimensional
cubic lattice ${\bf Z}^d$ (i.e.\ we set the lattice constant equal to 1).
In intermediate steps of our derivation we will implicitly assume that
$\Lambda_L$ is a finite cube (i.e.  $\vx=(x_1,\ldots,x_d)$ with
$-L/2\leq x_i< L/2$, $L$ a integer multiple of $4$ and $<\infty$) with a
finite number $L^d$ of points, but we will eventually take the
thermodynamic limit $L \to\infty$. We are mainly interested in $d=2$ and
$3$.

To define the Hubbard model in this setting, we introduce fields at every
point $x=(\tau,\vx)$ which are Grassmann variables $\bar\psi_\sigma(x)$,
$\psi_\sigma(x)$ carrying a spin index $\sigma=\uparrow,\downarrow$.
The action is  $S =S_2 + S_4$, with the quadratic
part $S_2 = -\left( {\bar\psi} , {G_0}^{-1} \psi\right)$; we use a
matrix notation i.e.\ $\left( f,g\right) =
\sum_{x,\sigma} f_{\sigma}(x) g_{\sigma}(x)$,
$\sum_x\equiv\int_0^{\beta}{d}\tau \sum_{\vx\in\Z^d}$, and
$G_0^{-1}(x-y)$ is the inverse of the free electron propagator
with the Fourier transform
\eq
\label{G0}
\left(G_0\right)^{-1}(k) = \left[ {i}\omega_n +\mu -\epsilon(\vk)
\right]\sigma_0
\eqend
where
\eq
\label{eps}
\epsilon(\vk) = -2t\sum_{i=1}^d  \cos(k_i)
\eqend
is the Hubbard band relation and $\sigma_0$ the $2\times 2$ (spin) unit
matrix. Here $\omega_n=(2n+1)\pi/\beta$, $n$ integer, are the usual electron
Matsubara frequencies, $\vk=(k_1,\ldots,k_d)$, $-\pi\leq k_i\leq \pi$, are
the pseudo momenta, and $\mu$ is the chemical potential.  The Hubbard
interaction term is $S_4 = U\sum_x \bar\psi_\uparrow(x)\psi_\uparrow(x)
\bar \psi_\downarrow(x) \psi_\downarrow(x)$. With that we can write the
partition function of the Hubbard model as $Z=\int {\cal D}\bar\psi {\cal
D}\psi\, e^{-S}$ where $\int {\cal D}\bar\psi {\cal D}\psi=\prod_{x,\sigma}
\int {d}\bar\psi_\sigma {d}\psi_\sigma$ are Grassmann integrals as usual
\cite{NO}.

We now come to an important technical point.  We recall that there are many
equivalent ways of writing the Hubbard interaction $S_4$.  This ambiguity
corresponds to an ambiguity in how to introduce Hubbard-Stratonovitch (HS)
fields, and it turns out that the saddle point equations one finally
obtains depends on the choice of HS fields \cite{Wolff}.  To obtain an
appropriate form for the partition function, we use ${\rm U}(1)$ (gauge) and
${\rm SU}(2)$ (spin rotation) invariance of the Hubbard model.  We note that
these symmetries are manifest for the term $S_2$, but $S_4$ as written
above is not explicitly ${\rm SU}(2)$ invariant.  To find a manifestly
symmetric
form for the partition function \cite{Schulz} we introduce
\eq
n = \sum_{\sigma}\bar\psi_\sigma \psi_\sigma , \quad
\vs =  \sum_{\sigma,\sigma'} \bar\psi_\sigma
(\vsigma)_{\sigma\sigma'}\psi_{\sigma'}
\eqend
which are the density and spin of the electrons; $\vsigma =
(\sigma_1,\sigma_2,\sigma_3)$ are the Pauli spin matrices as usual.
Manifestly invariant forms for the interaction would be $S_4=U\sum_x
n(x)^2/2 = -\sum_x \vs(x)^2/6$, but both of them do not make manifest
the Pauli principle: they contain terms $n_\sigma(x)^2 =
\left[\bar\psi_{\sigma}(x)
\psi_{\sigma}(x)\right]^2$ which are zero only by the Grassmann nature of the
fermion fields.  It is well-known that the latter is not preserved in
mean field theory.  In our framework this corresponds to
$n_\sigma(x)^2$ contributing non-zero terms to saddle point equations.
This is why both aforementioned choices for the interaction lead to
different saddle point equations, both different from the Hartree-Fock
equations, and are not appropriate.  We thus must make sure to have no such
terms present.  Writing $S_4=U\sum_x[n(x)^2-s_3 (x)^2]/4$ as above does
make the Pauli principle manifest [i.e.\ contains no terms
$n_\sigma(x)^2$], but so does
\eq
S_4= U\sum_x\frac{n(x)^2- [\ve(x)\cdot \vs(x)]^2}{4}
\eqend
with arbitrary unit vectors $\ve(x)$ [$\ve=(e_1,e_2,e_3)$].  In these
expressions, ${\rm SU}(2)$-invariance is not manifest.  However, we can average
over the directions $\ve(x)$ thus make ${\rm SU}(2)$ invariance and the Pauli
principle manifest at the same time.  Our Hubbard-Stratonovitch (HS)
transformation therefore is
\nonueqa
\exp\left[ -\frac{U}{4}\left( n^2- (\ve \cdot \vs)^2 \right) \right] =
\frac{1}{4\pi^2 U}\int_{\R} {d} \phi_0 \int_{\R}{d}\phi_s
\nonu\times \int_{4\pi}
{d}^2 \ve
\exp\left( -\frac{ \phi_0^2+\phi_s^2}{U} + {i} \phi_0 n +
\phi_s\, \ve\cdot \vs \right) \: .
\nonueqaend
(note that the l.h.s. here is independent of $\ve$).
We thus have, for each space-time point $x$, four HS fields $\phi_0$ and
$\vphi=\phi_s \ve = (\phi_1,\phi_2,\phi_3)$ corresponding to
density (charge) $n$ and spin $\vs$. Since $S_4$ is invariant under
$\ve(x)\to -\ve(x)$ we use $\int_{\R}{d}\phi_s\int_{4\pi}
{d}^2\ve = 2\int_{\R^3}{d}^3\vphi/\vphi^2$ and write the
interactions as
$$
e^{-S_4} = \int {\cal D}\phi \, e^{-\frac{1}{U} (\phi,\phi) -
(\bar\psi,\underline\phi \psi ) } \, ,
$$
where (up to an irrelevant constant factor which we drop)
\eq
\label{cD}
\int {\cal D}\phi =
\prod_{x}\int_{\R}{d}\phi_0
\int_{\R^3} \frac{{d}^3\vphi(x)}{\vphi(x)^2} \: .
\eqend
Here and in the following we use a convenient
matrix notation $(\phi,\phi)=\sum_{x,\alpha}\phi_{\alpha}(x)^2$
($\alpha=0,1,2,3$), and
\eq
\underline\phi(x) = {i}  \sigma_0\phi_0(x) +  \vsigma\cdot\vphi(x) \, .
\eqend
With that the electron can be integrated out and the partition function
becomes a path integral over the HS fields,
\eq
\label{Z}
Z= \int {\cal D}\phi \, e^{-\cF(\phi)}\, .
\eqend
The HS action is equal to
\eq
\label{F}
\cF(\phi) = \frac{1}{U} (\phi,\phi)  - {\rm Tr}\log(G_0^{-1} -
\underline\phi) \: .
\eqend
The trace ${\rm Tr}$ formally is defined for operators
$A=[A_{\sigma\sigma'}(x,y)]$ (acting on electron 1-particle states) as
${\rm Tr} A\equiv \sum_{x,\sigma} A_{\sigma\sigma}(x,x)$.

We note that all details of the form of this path integral are essential
for the approximations introduced in the next section to work.  For
example, the presence of the variable ${\bf e}(x)$ allows to describe, in a
natural way, the effect of changes in the direction of the AF order
parameter.

The free energy density (or rather thermodynamic potential) is defined as
\eq
\Omega(\beta,\mu) =-\lim_{L \to\infty}\frac{1}{L^d\beta } \log(Z)\, ,
\eqend
and the electron density fixes the chemical potential via the equation
\eq
\label{filling}
\xx = - \frac{\partial\Omega}{\partial\mu}\, .
\eqend
We find it convenient to make particle-hole symmetry manifest in our
formalism, and not use the electron density $n$ (average particle number
per site) but the doping parameter $\xx= n-1$.  Thus $-1\leq\xx\leq 1$, and
$\xx=0$ corresponds to half filling.  Then interchanging particles and
holes simply amounts to changing $\mu\to-\mu,\quad \xx\to-\xx$.
Technically this is achieved by defining
\eq
\label{reg1}
\frac{1}{\beta}{\sum_{\omega_n} } ^{ (reg)}
\frac{1}{{i}\omega_n-E} =
-\frac{1}{2}\tanh\left(\frac{\beta E}{2}\right) \equiv f_\beta(E).
\eqend
Here `$reg$' indicates that this sum is only conditionally convergent and
thus a certain summation prescription is used --- this is what is defined
by this formula.  This definition amounts to choosing a particle-hole
symmetric summation prescription for the conditionally converging Matsubara
sum defining the Fermi distribution function $f_\beta(E)$ [i.e.\ the
standard Fermi distribution function $(e^{\beta E}+1)^{-1}$ is shifted by
a constant so that it becomes odd under exchange $E\to -E$].  We stress
that this choice is just a matter of convenience: of course, one could
use the standard Fermi distribution function throughout at the cost of
having more complicated formulas for the particle$\leftrightarrow$hole
transformation.  We will also need the following Matsubara sum
\eq
\label{reg2}
\frac{1}{\beta}{\sum_{\omega_n} } ^{ (reg)}
\log\left( {i}\omega_n-E \right) =
\frac{1}{\beta}\log 2\cosh \Bigl(\frac{\beta E}{2}\Bigr) \equiv
\Ln(E)
\eqend
formally obtained by integrating Eq.\ (\ref{reg1}) [here the regularization
`$reg$' also requires to drop an infinite but $E$-independent
term; in the standard conventions the r.h.s. here would be
$\log(1+e^{-\beta E})/\beta$].

We finally give an exact formula for the 2 point Green function
$G=[G_{\sigma\sigma'}(x,y)]$ of the
Hubbard model,
\begin{mathletters}
\eq
G = \frac{1}{Z} \int {\cal D}\phi \, e^{-\cF(\phi)}\, G_0(\phi)\,
\eqend
where
\eq
\label{Dyson}
G_0(\phi)= [G_0^{-1} - \underline\phi]^{-1}
\eqend
\end{mathletters}is the Green function of non-interacting fermions in an
external field $\phi$.  (This formula is easily obtained from standard
expressions for fermion Green functions in the path integral formalism
\cite{NO}.) Similar formulas holds for all other Green functions.

\section{Restricted path integral method}
In this section we present the restricted path integral method and derive
our generalized mean field equation (\ref{Om}).  We first sum over the N\'eel
states since this immediately leads to Eq.\ (\ref{Om}).  To clarify the
meaning of this equation, we then show that it is also obtained if one sums
certain classes of states with local AF order but no long-range
correlations.  We finally outline how to systematically refine our
approximation.

\subsection{Summing over N\'eel states}

For a boson configuration Eq.\ (\ref{Neel}) describing AF long range order,
the HS action (\ref{F}) can be evaluated exactly using Fourier
transformation.  We obtain $\cF = \beta L^d \, f_0$ (i.e.\ the action is
proportional to the spacetime volume).  In the thermodynamic limit,
\eq
\label{f0}
f_0(r,s) = \frac{r^2+s^2}{U} - \intdk [\Ln(E_+) + \Ln(E_-)]
\eqend
with the function $\Ln(E)$ defined in Eq. (\ref{reg2}), and
\eq
\intdk  \equiv \int_{-\pi\leq k_i\leq \pi}\frac{{d}^d k}{(2\pi)^d}
\eqend
means integration over the Brillouin zone. Moreover,
\eq
\label{Epm}
E_\pm = {i} r-\mu \pm\sqrt{\epsilon(\vk)^2 + s^2}
\eqend
are the AF bands as usual.  Note that $f_0$ does not depend on the
direction $\ve$ of the AF order, as expected.  A formal derivation of
this result from Eq.\ (\ref{F}) can be found in Appendix~A.  It has a simple
physical interpretation: $-\intdk\Ln(E_\pm)$ is the free energy density of
non-interacting fermions with dispersion relations $E_\pm(\vk)$.

>From this we get the doping of the fermions in the
N\'eel background (\ref{Neel}) as $\xx(r,s) = -\partial f_0(r,s)/\partial
\mu$ i.e.\
\eq
\label{rhors}
\xx(r,s) = \intdk\left[ f_\beta(E_+) + f_\beta(E_-)  \right]
\eqend
where with $f_\beta(E)$ Eq.\ (\ref{reg1}) is our fermion distribution
function.

We now approximate the path integral (\ref{Z}) by summing only over the
N\'eel states Eq.\ (\ref{Neel}).  Since the integrand then depends only on
$r$ and $s$, the path integration $\int{\cal D}\phi$ Eq.\ (\ref{cD}) reduces to
$\int_0^\infty{d} s\int_{-\infty}^\infty{d} r$ (up to an irrelevant
constant; we introduced spherical coordinates,
$\int_{\R^3} {d}^3\vphi/\vphi^2 =
\int_0^\infty{d} s\int_{4\pi}{d}^2\ve$). Thus we obtain
\eq
\label{eq}
Z = \int_{-\infty}^\infty {d} r \,\int_0^\infty {d} s \,
e^{- V\, f_0(r,s) }
\eqend
where $V = \beta L^d$.  Using the definition $Z=\exp(-V\, \Omega)$ we
get Eq.\ (\ref{Om}).  With that, the particle number constraint
(\ref{filling}) becomes
\eq
\label{rho}
\xx = \frac{1}{Z} \int_{-\infty}^\infty {d} r \,\int_0^\infty {d} s \,
e^{- V\, f_0(r,s) }  \xx(r,s)
\eqend
with $\xx(r,s)$ defined in Eq.\ (\ref{rhors}).  We are interested in the
thermodynamic limit $V\to\infty$, thus the saddle point evaluation of
the integrals in Eqs.\ (\ref{eq}) and (\ref{rho}) is exact.  In case there is
only one relevant saddle point, we recover Hartree-Fock theory restricted
to the N\'eel states.

\subsection{Summing over states with local AF order}
We now show that Eq.\ (\ref{eq}) is also obtained if sum over configurations
with only local AF order, i.e.\ configurations of the form (\ref{Neel1})
where $r(x)$, $s(x)$ and $\ve(x)$ do not vary much in spacetime.
In the following we restrict ourselves to configurations $\phi$
such that $\cF(\phi)=\cF_0(\phi) + \cF_1(\phi)$ with $\cF_0$ Eq.\
(\ref{HSaction}) and $\cF_1$ negligible, i.e.\ $\lim_{L\to\infty}
\cF_1/\beta L^d =0 $; in this case we write $\cF\simeq
\cF_0$ . We do not attempt to further specify what
these configurations are.  The only point important for us is that there
are such configurations without long-range order.

Using Eq.\ (\ref{Neel1}), the path integration $\int {\cal D}\phi$ Eq.\
(\ref{cD}) becomes
$$
\prod_x \int_{-\infty}^\infty {d} r(x) \,\int_0^\infty {d} s(x)
\int_{4\pi}{d}^2\ve(x) \: .
$$ We see that if we sum over configurations $\phi$ for which $\cF\simeq
\cF_0$ Eq.\
(\ref{HSaction}), the integration over the $\ve(x)$ is irrelevant
(it contributes only a constant factor to $Z$ Eq.\ (\ref{Z}) which can be
dropped).

We first sum over all such configurations with $r(x)=r$ and $s(x)=s$
independent of $x$ but $\ve(x)$ changing (i.e.\ the magnitude of the AF
order parameter is constant but not its direction).  For all these
configurations we again obtain $\cF \simeq \beta L^d \, f_0(r,s)$, thus
summing over these configurations we obtain Eq.\ (\ref{eq}) as before.

We now sum over even larger classes of configurations allowing also for
spacetime dependent $r(x)$ and $s(x)$ which have sufficiently large spatial
and temporal correlations lengths $\ell_{space}$ and $\ell_{time}$,
respectively.  One can regard $\ell_{space}$ and $\ell_{time}$ as variation
parameters: calculate $Z$ Eq.\ (\ref{Z}) by summing over a class of
configurations characterized by these two parameters and determine them
such as to minimize the free energy $\Omega$.  In this case we also obtain,
in a good approximation, Eq.\ (\ref{eq}) but now with $V \approx
(\ell_{space})^d \ell_{time}$ the correlation volume (note that these
correlations only refer to the magnitude of the AF order but not to its
direction).  To illustrate that the precise structure of the boson
configurations summed over does not affect the final result, we consider
two different classes of configurations.  The first class contains
configurations $\phi(x)$ such that most points $x$ belong to a sufficiently
large region such that $\phi(x)$ restricted to this region equals a N\'eel
configuration Eq.\ (\ref{Neel}).  For configurations in the second class, the
Fourier modes of $r(x)$, $s(x)$ and ${\bf e}(x)$ have support sufficiently
close to $k=0$ but otherwise are arbitrary.

A configuration in the first class can be characterized as follows: set
$x=X+\xi$ where $\xi$ is a coordinate in a block [i.e.\ $s(X+\xi)$,
$r(X+\xi)$ are (approximately) independent of $\xi$], and $X$ labels
different blocks.  For these configurations $\cF(\phi)\simeq
\sum_x f_0\bigl(r(X),s(X)\bigr)$ which equals $\sum_X V_X
f_0\bigl(r(X),s(X)\bigr)$, and this defines the correlation volumes
$V_X$.  The correction $\cF_1$ to this should be proportional to the
number of points which belong to the boundaries of different blocks (whatever
their form is), thus it can be made negligibly small by choosing
$V_X$ sufficiently big.  We now sum over these configurations.  Since
$\int {\cal D}\phi=\prod_{X}\bigl(\prod_\xi
\int{d}\phi_0(X+\xi){d}^3\vphi(X+\xi)/\vphi(X+\xi)^2\bigr)$,
the path integral in Eq.\ (\ref{Z}) restricted to these configurations
gives
\[ Z = \prod_{X}
\int{d}\phi_0(X)\int \frac{{d}^3\vphi(X)}{\vphi(X)^2}
e^{-V_X f_0\bigl(r(X),s(X)\bigr)}
\]
which we can write as $ Z=\prod_X \exp(-V_X \Omega_X) $ where
$\Omega_X=\Omega(V_X)$ is defined by Eq.\ (\ref{eq}) with $V$ replaced by
$V_X$.  We thus get $
\Omega = \sum_X V_X \Omega_X/\sum_X V_X \: .
$ If we assume the configurations summed over such that $V_X=V$ is
independent of $X$, we immediately obtain Eq.\ (\ref{eq}).  More generally,
we also obtain Eq.\ (\ref{eq}) if $V_X$ is allowed to depend on $X$;
$V$ in this case is to be interpreted as an average correlation volume.

We now turn to configurations in the second class.  Let $s(k)$ and $r(k)$
be the Fourier transforms of $s(x)$ and $r(x)$ in Eq.\ (\ref{Neel1}).  We
consider boson configuration where these are different from zero only for
`small' $k$, i.e.\ $|\omega_n|<1/\ell_{time}$ and $|\vk|<1/\ell_{space}$.
For these we can use the approximation $\cF(\phi)\simeq
\sum_k' V f_0\bigl(r(k),s(k)\bigr)$ (which defines $V$) where the prime
indicates summation (or integration) over `small' momenta only.  If we
restrict the path integration in Eq.\ (\ref{Z}) to these configurations,
$\int
{\cal D}\phi$ becomes $\prod'_k \int{d}\phi_0(k){d}^3\vphi(k)/\vphi(k)^2$,
and we again obtain (\ref{eq}).

\subsection{Refinements: an outline}

As discussed, Eq.\ (\ref{Om}) is one simple of many possible approximation of
the partition function $Z$ by finite dimensional integrals.  The simplest
way to obtain Eq.\ (\ref{Om}) is by summing over the N\'eel states Eq.\
(\ref{Neel}).  The strategy for refining this approximation is
by the following natural generalization: make an ansatz for a set of boson
configurations Eq.\ (\ref{Neel1}) parameterized by $N$ real parameters
$\alpha_i$ such that with one such configuration all others related to it
by symmetry transformations are also in this set.  If all these
configurations are periodic (the period can be large), the HS action
configuration will be proportional to the total space-time volume $\beta
L^d$, and it depends on the $\alpha_i$.  Restricting the path integral in
Eq.\ (\ref{Z}) to these configurations we get an approximation of $Z$ by a
finite dimensional integral over the $\alpha_i$, and the saddle point
evaluation for this integral is exact.

The non-trivial task is to find a useful ansatz for the boson
configurations.  We suggest to use numeric studies of HF
theory \cite{walls,HartreeFock,VLLGB} as a guide.  One other criterion for
a `good' ansatz is, of course, that one can easily (the best would be
analytically) evaluate the HS action for the configurations considered.

\section{Numeric results}

In this section we present numeric results for the solution of our mean
field theory Eq.\ (\ref{Om}).  We discuss in detail the calculation for the 2
dimensional Hubbard model ($d=2$) and the parameters $U/t=10$ motivated by
HTSC \cite{Review}.  To demonstrate that the phase diagram in this
approximation is qualitatively
always the same, we also present results for 3 dimensions ($d=3$ and
$U/t=10$) and weaker coupling ($U/t=2$, $d=2$).

Standard integration routines were used to numerically evaluate the
$\vk$-integrals defining the functions in Eqs.\ (\ref{f0}), (\ref{rhors}),
solve Eq.\ (\ref{r}) below etc.  In all our calculations we ensured that
numeric errors are negligible.

\subsection{Saddle point evaluation}

We now describe in detail how to evaluate the integral in Eq.\ (\ref{Om}) in
the limit $L\to\infty$ under the constraint Eq.\ (\ref{rho}).  We first
perform the $r$-integral, ($\mu$-dependence suppressed)
\eq
\label{f}
e^{-V f^*(s)} \simeq \int_{-\infty}^\infty{d} r\, e^{-V f_0(r,s) }
\qquad (V\to\infty)
\eqend
where $V=\beta L^d$ and $A \simeq B$ here means that
$\lim_{V\to\infty} [\log(A)-\log(B)]/V = 0$. The saddle point equation
for this integral is $\partial f_0/\partial r=0$, i.e.\
\eq
\label{r}
r =  - \frac{{i}}{2} U \xx(r,s)
\eqend
[we used Eq.\ (\ref{rhors})].  This equation has a unique purely imaginary
solution $r=r^*(s)$, and one can show that this saddle point dominates the
integral in Eq.\ (\ref{f}), and a standard saddle point evaluation gives
\eq
\label{fstar}
f^*(s)=f_0(r^*(s),s) \: .
\eqend
The complete justification of this result is
somewhat technical and deferred to Appendix~\ref{B1}.

With that we get from Eq.\ (\ref{eq})
\eq
\label{sdl}
Z= e^{-V \Omega} \simeq \int_0^\infty {d} s\, e^{-V f^*(s)}
\qquad (V\to\infty)
\eqend
Note that $r^*(s)$ enters here only in the combination $\tilde\mu=\mu-{i}
r^*(s)=\mu-U\xx(r^*(s),s)/2$.  Physically this can be naturally interpreted
as renormalization of the chemical potential by the Coulomb energy.  Since
$f^*(s)$ is real-valued, this integral is determined by the absolute
minimum (minima) of this function. Plotting the function $f^*(s)$
for different values of the chemical potentials $\mu$ thus is a simple
way to understand why degenerate saddle points occur.

\begin{figure}
\epsfxsize=9truecm\epsffile{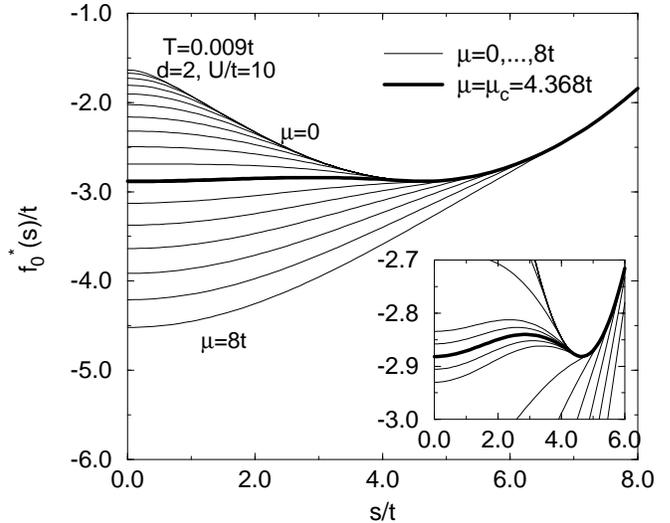}
\caption{Dependence of effective action $f^*$ in
Eq.~(\protect{\ref{fstar}}) on the magnitude of the antiferromagnetic order 
$s$ for a sequence of different values of the chemical potential and a 
temperature close to zero ($T=0.009t$) {\protect\cite{temp}}.  Parameters 
here are $U/t=10$ and $d=2$.  Note that there are sometimes two minima, and 
these become degenerate at a critical value of the chemical potential, 
$\mu=\mu_c$ (fat curve).  Inset: zoom with additional curves for 
$\mu=\mu_c(1+i/100)$, $i=0,\pm 1,\pm 2$.  }
\end{figure}

\begin{figure*}
\hbox{
\epsfxsize=9truecm\epsffile{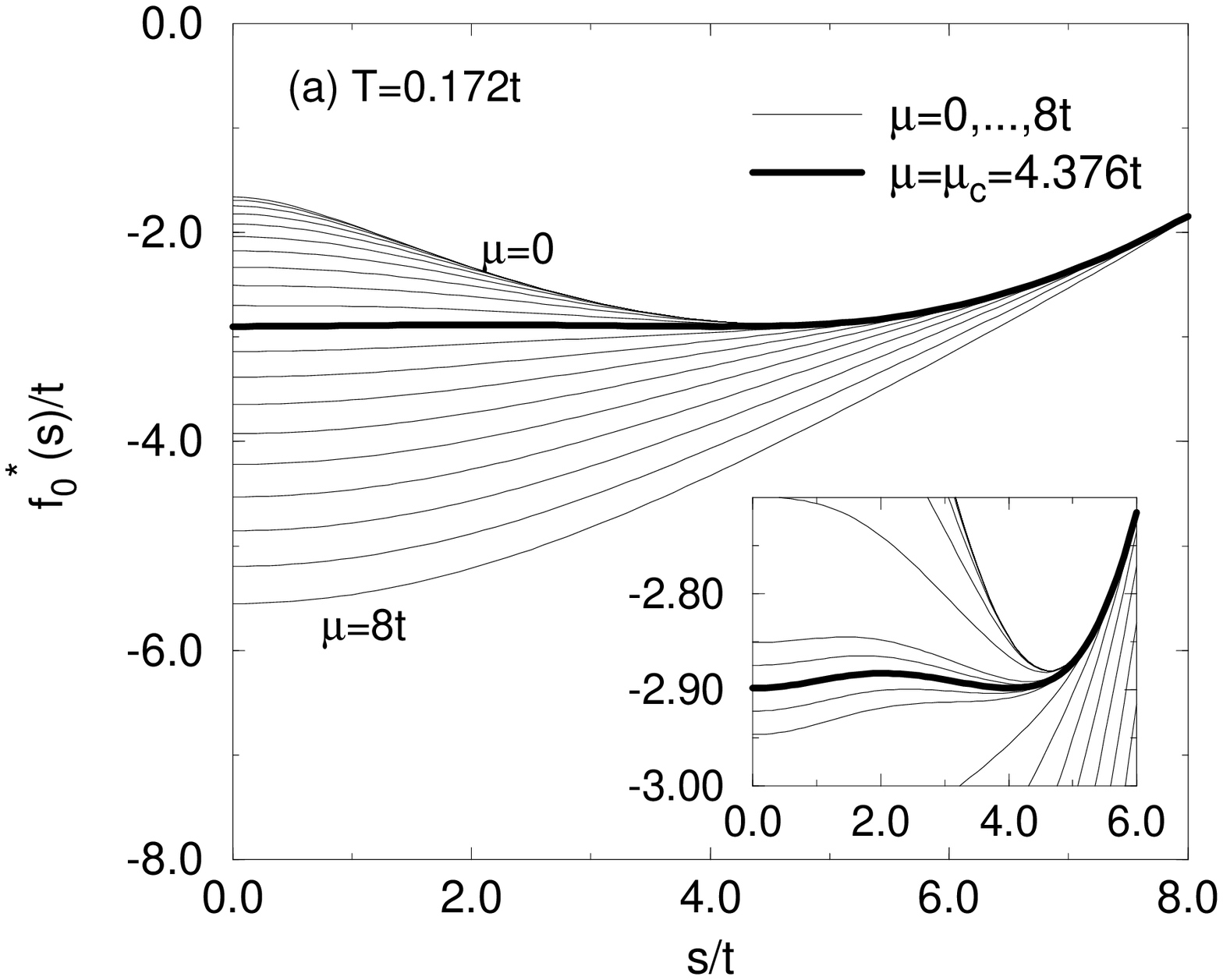}
\epsfxsize=9truecm\epsffile{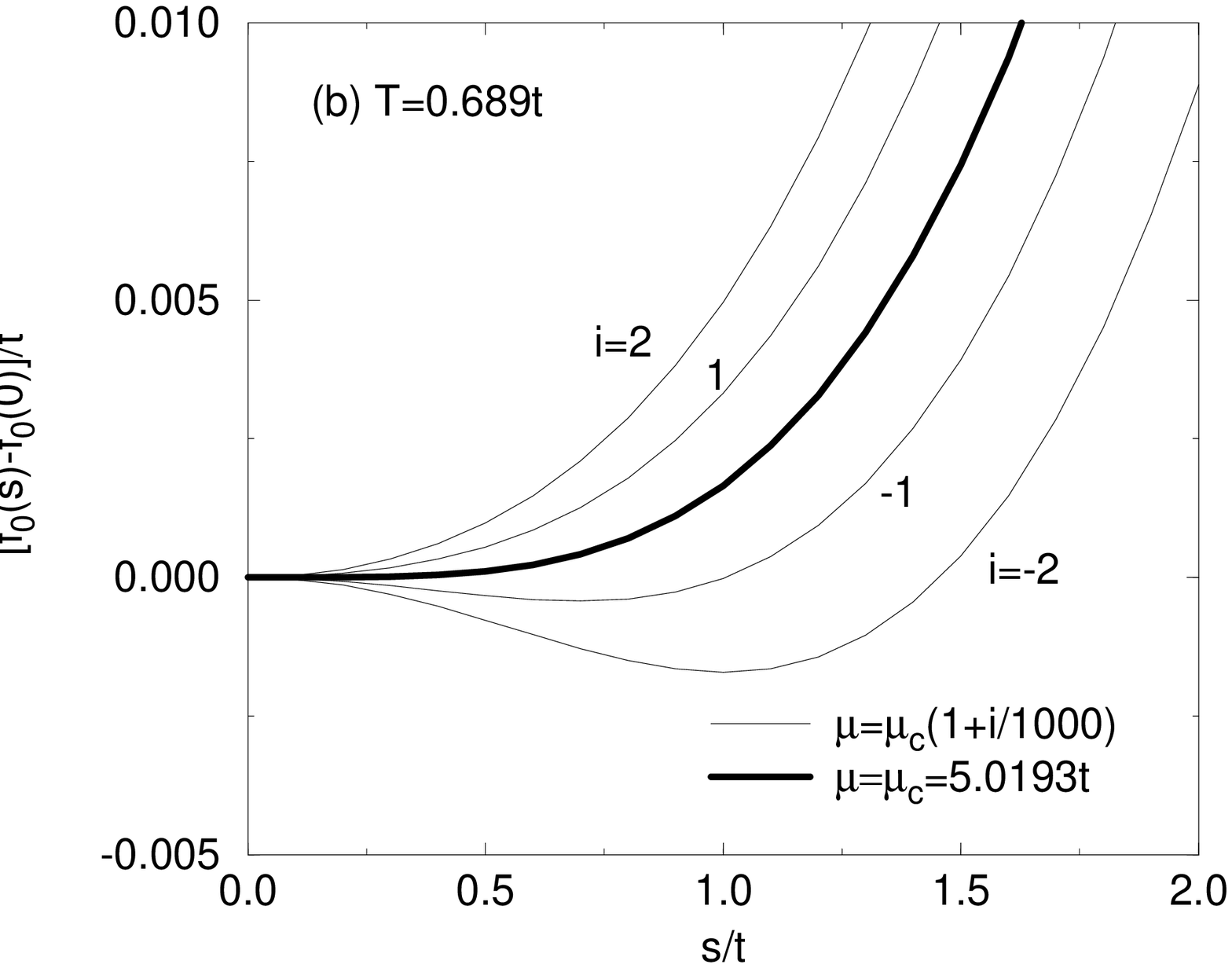}
}
\caption{Same as Fig.~1 but for different temperatures $T$ {\protect\cite{temp}}.  
(a): Higher temperature but still below $T^{\star}$.  Otherwise the same as 
in Fig.~1.  Note that the non-trivial minimum now becomes slightly 
$\mu$-dependent close to $\mu=\mu_c$.  (b): Temperature above $T^{\star}$.  
Note that there are no longer two minima possible at the same $\mu$.  We 
zoomed in close to $\mu=\mu_c$ where the non-trivial minimum merges with 
the trivial one and disappears.}
\end{figure*}

Fig.~1 shows such a plot at a fixed temperature close to zero
\cite{temp}.  We see that for $\mu=0$ up to some critical value $\mu=\mu_c$,
the absolute minimum of $f^*(s)$ is at a finite value $s=s^*$ which thus
dominates the integral in Eq.  (\ref{sdl}), $\Omega = f^*(s^*)$.  For
$\mu>\mu_c$ the minimum at $s=0$ takes over, and $\Omega = f^*(0)$.  We
refer to $s=s^*$ and $s=0$ as the non-trivial and trivial minima (or saddle
points), respectively.  We see from Fig.~1 that the nontrivial minimum
is always at the same value $s=s^*$ (independent of $\mu<\mu_c$), and also
$f^*(s^*)$ does not change with $\mu$.  Thus $\xx = -\partial
f^*(s^*)/\partial\mu$ is always 0, and the nontrivial minimum can only
account for half filling.  The physical interpretation of this is that for
$s=s^*>0$, we have AF bands with a gap, and as long as $\mu$ is in this gap
it cannot affect the doping.  For $\mu>\mu_c$, we see that $\xx=-\partial
f^*(0)/\partial\mu$ is always larger than $\xx_2=-\partial
f^*(0)/\partial\mu|_{\mu=\mu_c}$ (which is larger than $0.5$ here; note
that $\xx$ is monotonically increasing with $\mu$).  Thus the only doping
which can be obtained with the non-trivial minimum $s=s^*$ is $\xx=0$, and
for the trivial minimum $s=0$ only dopings $\xx>\xx_2$ are possible.  A
finite doping regime $\xx_1<\xx<\xx_2$ is left out by this ($\xx_1=0$ here
but will be non-zero for higher temperatures; see below).  {\em The only
way to get doping in this regime is to have $\mu$ so close to $\mu_c$ that
both saddle points can contribute to the integral Eq.\ (\ref{sdl})}.  Indeed,
for $\mu=\mu_c+\delta\mu$ we have $f^*(s^*)-f^*(0) =
-(\xx_1-\xx_2)\delta\mu+{\cal O}(\delta\mu^2)$ ($\xx_{1,2}=-\partial
f^*(s)/\partial\mu|_{\mu=\mu_c}$ at $s=s^*$ and $0$, respectively), which
suggests that we can adjust $\delta\mu={\cal O}(1/V)$ such that the saddle
points $s=s^*$ and $s=0$ contribute with relative weight $w$ and $(1-w)$ to
the partition function Eq.\ (\ref{sdl}) and dopings $\xx$ between $\xx_1$ and
$\xx_2$ are possible,
\eq
\label{weight}
\xx = w\, \xx_1+ (1-w)\, \xx_2 \: .
\eqend
(We give a more careful argument in Appendix~\ref{B2}.)  This equation
fixes $w$ by doping, and one can now forget about $\delta\mu$ which, for
$V\to\infty$, becomes zero.

Standard mean field theory corresponds to evaluating the integral (\ref{eq})
insisting on having only one relevant saddle point.  Then
$\Omega=f_0(r,s)$ where $r$ and $s$ solve the saddle point equations
$\partial f_0/\partial r = \partial f_0/\partial s=0$, i.e.\
Eqs.\ (\ref{r}) and
\eq
\label{s}
s = - s\, U\intdk \left[\frac{f_\beta(E_+)-f_\beta(E_-)}{E_+-E_-}\right] \: .
\eqend
The doping constraint Eq.\ (\ref{filling}) becomes
\eq
\label{rh}
\xx=\xx(r,s)
\eqend
which makes Eq.\ (\ref{r}) trivial, $r=-{i} U\xx/2$, and one needs to consider
only Eqs.\ (\ref{s}) and (\ref{rh}): for fixed $s$, Eq.\ (\ref{rh}) fixes the
renormalized chemical potential $\tilde\mu=\mu-{i} r = \mu -U\xx/2$, and Eq.\
(\ref{s}) determines the possible $s$-values.  In general one gets more
solutions.  However, from our discussion above it is clear that in the
doping regime $\xx_1<\xx<\xx_2$ none of these solutions is appropriate: the
trivial solution $s=0$ is not an absolute minimum of $f^*(s)$, and the
non-trivial solution $s\neq 0$ corresponds to the maximum of $f^*(s)$.
We conclude that {\em the correct mean field theory corresponds to a
saddle point evaluation of the integral (\ref{Om}) under the constraint
(\ref{filling}) as described above: one can have two relevant saddle points
of equal importance.}

Up to now our discussion was restricted to low temperatures.  At higher
temperatures the situation is similar, only that the mixed phase occurs in
a doping region $\xx_1<\xx <\xx_2$ where $\xx_1>0$, and for dopings $0 \leq
\xx \leq \xx_1$ one can have the nontrivial saddle point alone.
This can be easily understood from the plot of $f^*(s)$ for different
chemical potentials $\mu$ in Fig.~2(a): The nontrivial minimum $f^*(s^*)$ is
completely independent of $\mu$ again, except in a tiny $\mu$-region close
to the critical $\mu_c$ where the two minima become degenerate [inset of
Fig.~2(a)].  The explanation for this is simple: as long as $\mu$ is in the AF
gap, it has no effect on the doping, only when it comes close to the band
edge the doping starts to depend on $\mu$ due to thermal effects (i.e.\
since Fermi distribution function no longer is a step function).
Increasing $\mu$ further, the nontrivial minimum starts to move but soon
the trivial minimum takes over.  The value $\xx_1=-\partial
f^*(s^*)/\partial\mu|_{\mu=\mu_c}$ is the upper limit for the doping which
can be obtained by the nontrivial minimum without superposing the trivial
one.

The phase boundary $\xx_1$ of the pure AF phase increases with temperature
and approaches the boundary $\xx_2$ of the free phase which is quite
temperature independent [see Fig.~3].  At some temperature $T=\Tst$,
$\xx_1=\xx_2$, and the mixed phase disappears.  For temperatures $T>\Tst$,
there is a direct second order phase transition at the critical filling
$\xx_2$ from the AF to the free phase with no mixed phase between.
Fig.~2(b) shows a typical plot of $f^*(s)-f^*(0)$ for different chemical
potentials $\mu$ for such high temperature case.  We restricted ourselves
to the interesting region close to the critical $\mu=\mu_c$ defining
$\xx_2$.  One sees that for $\mu < \mu_c$, $f^*(s)$ has a nontrivial
minimum $s^*(\mu)$ lower than the trivial one, but this minimum approaches
the trivial with increasing $\mu$ and merges with it at $\mu=\mu_c$ [i.e.\
$s^*(\mu_c)=0$].  The doping $\xx_2$ is equal to $-\partial
f^*(0)/\partial\mu|_{\mu=\mu_c}$.  The qualitative difference to Figs.~1
and 2(b) is that there exists no $\mu$ where the two minima are degenerate.

Our discussion above suggests that occurrence of degenerate saddle points is
a rather stable phenomenon and should survive corrections to mean field
theory, e.g.\ by fluctuations: A non-trivial saddle point can dominate the
partition function only in a finite $\mu$-regime, and whenever not all
doping values are realized in that regime degenerate saddle points occur.
For the same reason we expect that degenerate saddle points are typical
also for other interacting lattice fermion models.

\subsection{Phase diagrams}

\begin{figure}
\epsfxsize=9truecm\epsffile{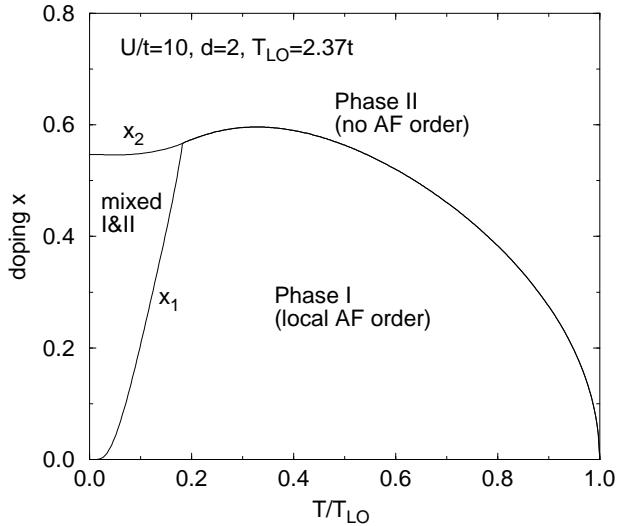}
\caption{Phase diagram for the 2D Hubbard model and parameters $U/t=10$. 
At fixed low temperature $T$ there is a homogeneous Phase I with local AF
order for doping less than ${\rm x}_1$, and a homogeneous metallic Phase II
without AF correlations for doping larger than ${\rm x}_2$.  In the regime
between ${\rm x}_1$ and ${\rm x}_2$ these two phases coexist.  ${\rm
x}_1=0$ for low temperatures, and it increases with $T$ until a
characteristic temperature $T^{\star}$ where it merges with ${\rm x}_2$ and
the mixed phase disappears.  For higher $T$ there is a smooth transition
from Phase I to Phase II.  $T_{LO}$ is the highest temperature at which
{\em local} AF order exists.  }
\end{figure}

In the last subsection we explained how our generalized mean field theory
allows to determine the different regimes of the Hubbard model.  In
Fig.~3 we have plotted the resulting phase diagram.  We see the 3
regions, the pure AF region with only the nontrivial saddle point, the free
region with the trivial saddle point only, and the mixed region where both
saddle points are relevant.  The temperature $T_{LO}$ is the largest
temperature where a nontrivial saddle point can contribute, and the other
characteristic temperature $\Tst$ is the upper limit for the mixed region
to exist, as discussed.  We note that $T_{LO}$ is equal to the highest
temperature where the HF equations restricted to N\'eel states Eq.\
(\ref{Neel}) at half filling have a non-trivial solution. Usually this
temperature is interpreted as the N\'eel temperature.  However, from our
derivations it is clear that we should interpret this temperature as the
highest temperature where {\em local} AF correlations occur, and this is to
be expected higher than the N\'eel temperature where long-range order
disappears.

\begin{figure*}
\hbox{
\epsfxsize=9truecm\epsffile{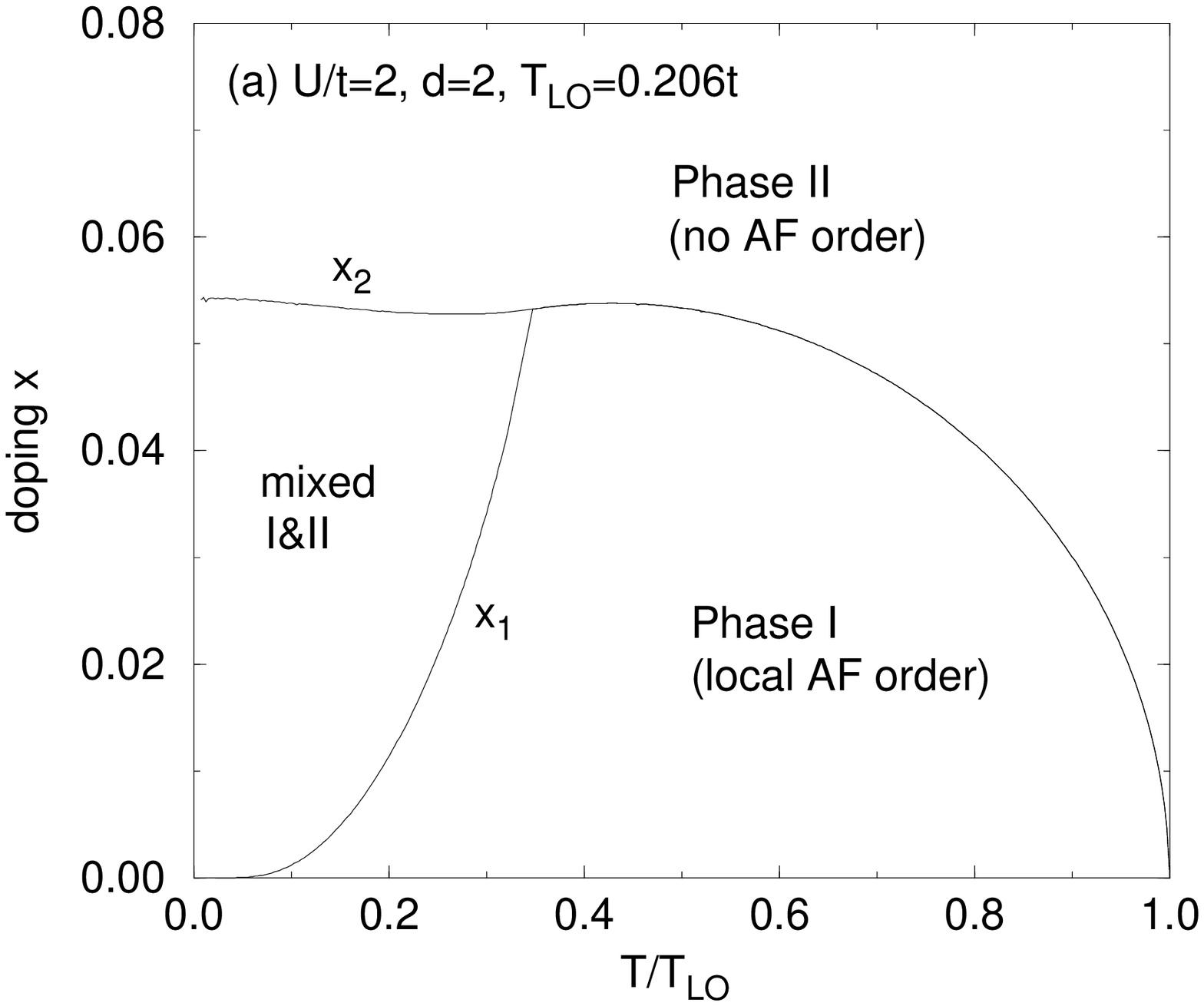}
\epsfxsize=9truecm\epsffile{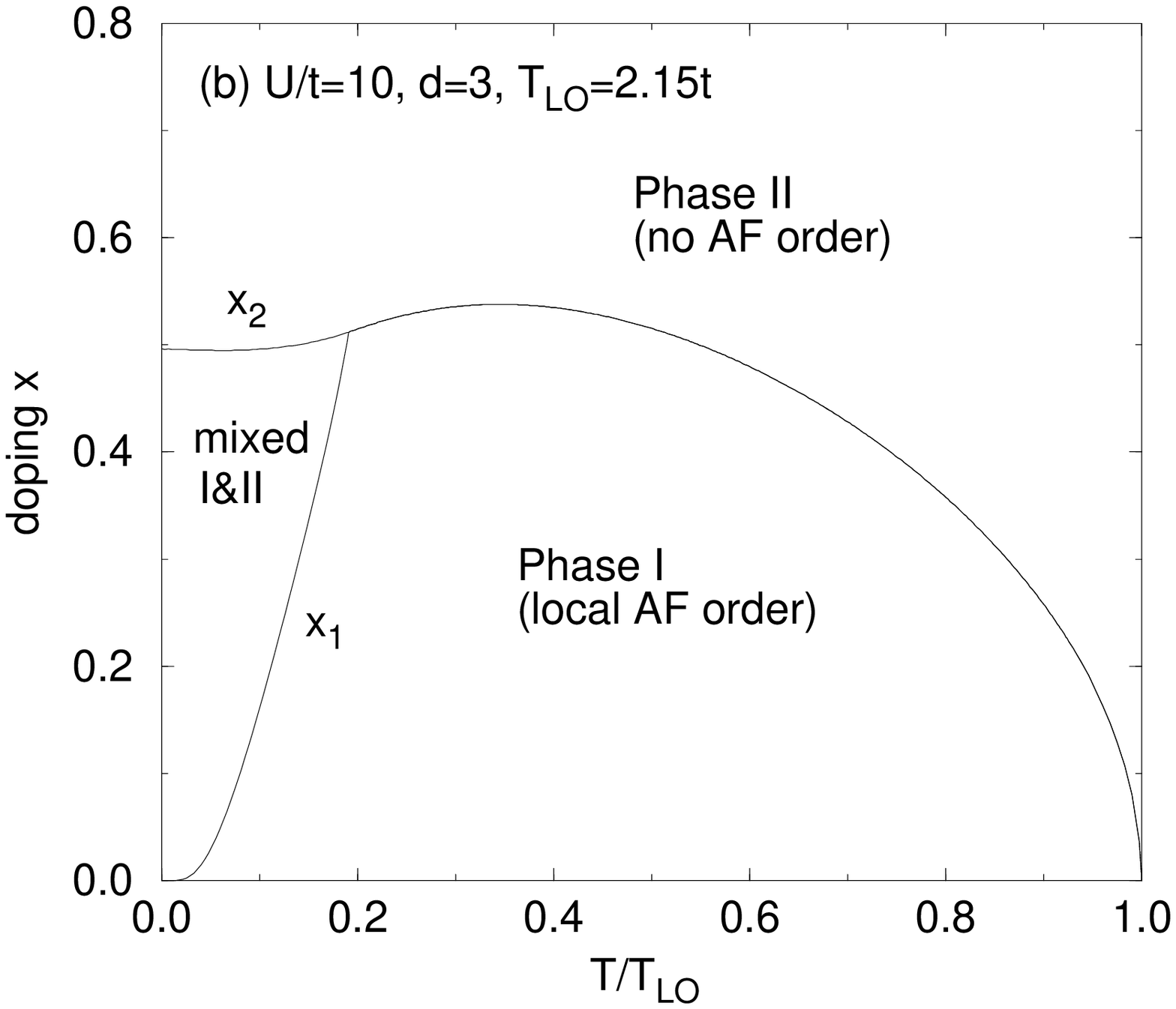}
}
\caption{Phase diagram for the Hubbard model and different parameters.
(a): $U/t=2$ and $d=2$, (b): $U/t=10$ and $d=3$.
Note that the phase diagrams are qualitatively the same as the one 
in Fig.~3.}
\end{figure*}

Our detailed discussion above was for the Hubbard parameters $U/t=10$ and
$d=2$.  However, the results obtained are representative.  Do illustrate this
we give two further examples for phase diagrams in Fig.~4(a) ($U/t=2$, $d=3$)
and Fig.~4(b) ($U/t=2$, $d=2$).

\section{Fluctuations}
In this section discuss the effect of fluctuations.  We first outline the
general scheme how to include fluctuations in our formalism.  As an
illustration, we discuss in more detail the effect of fluctuations in the
direction of the AF order using a simple approximation.

\subsection{General formalism}
Our derivation of mean field theory in Section III suggests that a simple
method to take into account fluctuations is as follows: We consider boson
configurations of the form $\phi=\phi^{(0)} + \delta\phi$ where
$\phi^{(0)}$ are the `little' varying ones leading to (generalized) mean
field theory, and $\delta\phi$ correspond to `fast' fluctuations.  We
assume the latter are such that we can expand
\begin{mathletters}
\eqa
\label{gauss}
\cF(\phi^{(0)} + \delta\phi) - \cF(\phi^{(0)}) \simeq
 \sum_{x,\alpha} B_\alpha(x)\delta\phi_\alpha(x) +\nonu
\sum_{x,y,\alpha,\beta}
C_{\alpha\beta}(x,y)\delta\phi_\alpha(x)\delta\phi_\beta(y)
\eqaend
and neglect the ${\cal O}(\delta\phi^3)$-terms.
The coefficients here are
\eqa
B_\alpha(x)&=&\left. \frac{ \delta
\cF(\phi)}{\delta\phi_{\alpha}(x)}\right|_{\phi=\phi^{(0)}}\nonu
C_{\alpha\beta}(x,y) &=& \left. \frac{1}{2}
\frac{ \delta^2 \cF(\phi)}{\delta\phi_{\alpha}(x)\delta\phi_{\beta}(y)}
\right|_{\phi=\phi^{(0)}}
\eqaend
and depend only on $\phi^{(0)}$.
We thus get a simple framework for taking into account AF- and charge
fluctuations $\delta\vphi(x)$ and $\delta\phi_0(x)$ around mean field
theory: integrating over the fluctuations $\delta\phi_\alpha(x)$
gives a correction term $\delta\cF(\phi^{(0)})$ to $\cF(\phi^{(0)})$
which describes the effect of fluctuations.  Summing then over the
$\phi^{(0)}$ leads to a finite dimensional integral as before.
Symbolically we can summarize this procedure as follows,
\eqa
e^{-\delta\cF(\phi^{(0)}) } &=& \int_{\delta\phi(x) }
e^{-\cF ( \phi^{(0)} +\delta\phi ) + \cF(\phi^{(0)})} \nonu
Z &=& \int_{ \phi^{(0)}(x) } e^{-\cF(\phi^{(0)} ) - \delta\cF(\phi^{(0)}) }
\eqaend
\end{mathletters}There is one important point to keep in mind: one can
integrate only over fluctuations $\delta\phi$ where the second term in Eq.\
(\ref{gauss}) is positive definite, and this is an important restriction on
which fluctuations can be taken into account in that way \cite{manfred}.
Selecting appropriate `fast' fluctuations therefore is quite delicate.
Also the summation over these fluctuations will be, in practice, a
non-trivial task and require further approximations.

\subsection{Example}
As an illustration we consider
fluctuations $\delta{\bf e}(x)$ around states $\phi^{(0)}$ defined in Eq.\
(\ref{Neel1}) with $r(x)=r$, $s(x)=s$ independent of $x$ and ${\bf
e}(x)={\bf e}^{(0)}(x)$ varying `not much'.  Then Eq.\ (\ref{gauss}) gives
\eqa
\cF(\phi) - \cF(\phi^{(0)}) =\beta L^d \,
\frac{\partial f_0(r,s)}{\partial s} s\,  {\bf e}^{(0)}(x)\cdot \delta{\bf
e}(x) \nonu+  {\cal O}(\delta\phi^2)\qquad\quad
\eqaend
where we restrict ourselves to the lowest-order correction term for
simplicity.  In a simple approximation, integrating out such fluctuations
amounts to replacing ${\bf e}^{(0)}(x)\cdot \delta{\bf e}(x)$ by its
average which should be an $x$-independent constant.  Note that this
constant depends on $r$, $s$, $\beta$, $\mu$, and on the class of
fluctuations considered.  In any case, since $[{\bf e}^{(0)}(x)+ \delta{\bf
e}(x)]^2 = {\bf e}^{(0)}(x)^2=1$, this constant is equal to the average of
$-\delta{\bf e}(x)^2/2$ and thus negative.  We get $\delta\cF =
\beta L^d\, f_1$  with
\eqa
f_1(r,s) = \beta L^d\, \frac{\partial f_0(r,s)}{\partial s} s\,  \alpha\,
, \quad
\alpha =  - \frac{1}{2}\left<\delta{\bf e}(x)^2\right>
\eqaend
[the averaging $<\cdot>$ of course defined through
$$
e^{-\delta\cF} = \int\, e^{-\cF(\phi) + \cF(\phi^{(0)})}
$$
where $\int$ here symbolizes integration (summation) over all
configurations ${\bf e}^{(0)}(x)$ and $\delta{\bf e}(x)$
taken into account]. This then leads to Eq.\ (\ref{Om})
with $f_0$ replaced by $f_0+f_1$.

To evaluate $\alpha$ is difficult.  However, one can easily estimate the
effect of these fluctuations by taking $\alpha$ as a fixed constant.  Then
one can determine the phase diagram as function of $\alpha$ similarly as
described in Section IV.  We do not present numeric results here and only
note that, even though $\alpha$ has quite a {\em quantitative} effect on the
phase diagram, the {\em qualitative} features of phase diagram are very
stable.

\section{Conclusions}
In this paper we presented a novel method for finding approximations for
interacting fermions models.  These approximations not only allow to
determine phase diagrams but also Green functions (i.e.\ observables) of
the systems.  Our discussion was for antiferromagnetic (AF) correlations
and the Hubbard model but is straightforward to extend to other models.  In
simple cases our approximations corresponds to generalized mean field
theory which allows the possibility of degenerate saddle points.
Occurrence of degenerate saddle points can be easily detected in our
formulation and reveals a frustration usually referred to as phase
separation.  However, our interpretation of this phenomenon is somewhat
more general: our approach allows to clarify the meaning of the mean field
approximation and shows that degenerate saddle points only mean that no
homogeneous mean field solution exists.  A more detailed description of the
systems is then necessary, and this requires a refined approximation taking
into account inhomogeneous states.

Our approach formalizes the simple and intuitive interpretation of mean
field theory as an approximating to the exact partition function by
restricting the summation over the huge set of states to a (small) subset
which, on physical grounds, are expected to represent the `most important
states'.  This makes manifest that mean field theory is not necessarily a
weak-coupling approximation, and it also makes manifest that the simplest
ansatz of translation invariant Hartree-Fock (HF) solutions not only
describes long-range ordered states but also states with only local
correlation.  Thus a non-trivial solution of HF equations restricted to
long-range ordered states should not be interpreted as long-range order but
only as existence of {\em local} correlations.  Moreover, our approach
gives a systematic way to find increasingly refined (and increasingly
complicated) mean field theories which give a homogeneous description of
the system even if the HF solutions are not translation invariant.  Our
approach also allows to include fluctuations in a natural way (the latter
is only outlined in this paper).

Many successful theories in solid state physics are based on simple mean
field theory i.e.\ HF equations and an ansatz allowing only for homogeneous
solutions.  If this latter ansatz is not appropriate, the system will be
much more complicated and not resemble a free fermion system.  This is the
case for the Hubbard model and probably most correlated fermions models.
This is one reason (and we believe: the reason) why our understanding of
these systems is still rather poor.  However, in such situation a
generalized mean field theory as proposed in this paper is still possible,
and we believe that this could be the starting point for a successful
theory of correlated fermions systems.\\

\noindent
{\bf Acknowledgments.} We thank P.\ Hedeg{\aa}rd, J.\ Lidmar, M.\
Salmhofer, and A.\ Sudb{\o} for helpful discussions.  This work was
supported by a grant from the Swedish Natural Science Research Council.

\begin{appendix}

\section{Free energy in N\'eel state}

In this appendix we evaluate the HS action Eq.\ (\ref{F}) for the N\'eel
configuration Eq.\ (\ref{Neel}) and thus prove Eq.\ (\ref{f0}).  The
non-trivial step obviously is to evaluate ${\rm Tr}\log (G_0
-\underline{\phi})$ for $\underline{\phi} = {i} r\sigma_0 +
s\,\ve\cdot\vsigma\,e^{{i}\vQ\cdot\vx }$.  We recall that our space
$\Lambda_L$ is a cube with points $\vx = (x_1,\ldots, x_d)$, $x_i$ integers
with $-L/2\leq x_i < L/2$.  We find it convenient to write points in
$\Lambda_L$ as $2\vx + \vn$ where $\vx\in\Lambda_{L/2}$ and
$\vn=(n_1,\ldots, n_d)$ with $n_j=0,1$.  Then the Green function
$G=(G_0^{-1} - \underline{\phi} )^{-1}$ is represented by a translation
invariant $2^{d+1}\times 2^{d+1}$-matrix,
\eqa
(G)_{\sigma\sigma' m_1n_1\ldots m_d n_d}(\tau-\tau',2\vx-2\vy) =\nonu
G_{\sigma\sigma'}(\tau,2\vx + \vm,\tau', 2\vy + \vn ) \, ,
\eqaend
and therefore,
\eqa
\label{A2}
{\rm Tr}\log G^{-1} = \nonu
\int_0^\beta {d}\tau
\sum_{\vx\in\Lambda_{L/2} }
{\rm tr} (\log G^{-1})(\tau-\tau,2\vx-2\vx) \nonu
=  \sum_{\vk\in \Lambda_{\pi/2}^* }
{\sum_{\omega_n} }^{(reg)}
{\rm tr} \log G^{-1}({i}\omega_n,\vk)
\eqaend
where $ {\rm tr}$ is the trace over the $2^{d+1}\times 2^{d+1}$-matrix
indices, $\Lambda_{\pi/2}^*$ is the set of all $\vk=(k_1,\ldots, k_d)$
with $k_i=2\pi\nu_i/L$ with $\nu_i$ integer and such that $-\pi/2\leq
k_i<\pi/2$; the Fourier transform is defined as $G(\vk) =
\sum_{\vx\in \Lambda_{L/2} } e^{-{i} \vk\cdot2\vx} G(2\vx)$.  To
evaluate Eq.\ (\ref{A2}), we first note that the matrix notation we use here
comes from representing fermion field operators as
\[
\psi_{\sigma n_1\ldots n_d}(\tau, 2\vx) =
\psi_\sigma (\tau, 2\vx + \vn) \, .
\]
>From this we see that $\underline{\phi} = {i} r\sigma_0 +
s\,\ve\cdot\vsigma\,e^{{i}\vQ\cdot\vx }$ is represented by
\[
{i} r\, \sigma_0\otimes\underbrace{\sigma_0\otimes \cdots
\otimes\sigma_0}_{\mbox{$d$ times}} + s\,\ve\cdot\vsigma\,\otimes
\underbrace{\sigma_3\otimes \cdots \otimes\sigma_3}_{\mbox{$d$ times}}
\]
(we use an obvious tensor notation).
We also need the hopping operators $(T_j\psi)_\sigma(\vx) =
t[\psi_\sigma(\vx+\ve_j) + \psi_\sigma(\vx -\ve_j)]$
where $\ve_j$ is the lattice unit vector in $j$-direction [i.e.\
$\ve_1=(1,0,\ldots,0)$ etc.]. It is easy to see that these
are represented by
\eq
T_j = \sigma_0\otimes \underbrace{\sigma_0\otimes \cdots
\otimes\sigma_0}_{\mbox{$j-1$ times}}
\otimes\, T(k_j)\otimes\underbrace{\sigma_0\otimes \cdots
\otimes\sigma_0}_{\mbox{$d-j$ times}}
\eqend
with
\eq
T(k_j) = -2t\cos(k_j)\left(\bma{cc} 0& e^{{i} k_j}\\  e^{-{i} k_j}
& 0 \ema \right)
\eqend
[$\vk=(k_1,\ldots,k_d)$]. Thus
\begin{mathletters}
\eq
G^{-1}(k) = ({i}\omega_n +\mu -{i} r)\,{\bf 1} -M
\eqend
where ${\bf 1}= \sigma_0\otimes\cdots\otimes\sigma_0$ is the $2^{d+1}\times
2^{d+1}$ unit
matrix and
\eq
M= s\,\ve\cdot\vsigma\,\otimes
\sigma_3\otimes\cdots\otimes\sigma_3 + \sum_{j=1}^d T_j \, .
\eqend
\end{mathletters}
We now evaluate $M^2$ and obtain
\[
M^2 = \bigl( s^2 + \sum_j \epsilon_j^2 \bigr){\bf 1} + \sum_{i<j} 2 T_i T_j
\]
where $\epsilon_j = -2t\cos(k_j)$ [we used $\sigma_3T(k_j)+T(k_j)\sigma_3=0$].
It is easy to diagonalize $M^2$: Let $U_j$ be the $2\times 2$-matrix
diagonalizing $T_j$, $U_j^{-1} T_j U_j =
\epsilon_j\sigma_3$, then $U=\sigma_0\otimes U_1\otimes\ldots\otimes U_d$
diagonalizes $M^2$,
\eqa
\bigl(U^{-1}M^2 U\bigr)_{\sigma\sigma'm_1n_2\ldots m_dn_d} =
\delta_{\sigma\sigma'}\delta_{m_1n_1} \cdots \delta_{m_dn_d}\nonu
\times
\Bigl(s^2+\sum_j\epsilon_j^2+2\sum_{i<j}\epsilon_i\epsilon_j(-)^{n_i+n_j}
\Bigr)\, .
\eqaend
Since $\epsilon_j(-)^{n_j} = -2t\cos(k_j+n_j\pi)$ for $n_j=0,1$ we see that
the eigenvalues of $M^2$ are ${s^2+\epsilon(\vk+\vn\pi)^2}$ where
$\epsilon(\vk)$ is the Hubbard band relation Eq.\ (\ref{eps}).  Moreover, for
$\ve'$ a unit vector orthogonal to $\ve$, the self-adjoint matrix
$C=\ve'\cdot\vsigma\otimes\sigma_3\cdots\otimes\sigma_3$ obeys $C^2={\bf
1}$ and $CMC=-M$.  We conclude that the eigenvalues of $M$ are $\pm
\sqrt{s^2+\epsilon(\vk+\vn\pi)^2}$, thus the eigenvalues of $\log
G^{-1}(k)$ are $\log\bigl({i}\omega_n-E_\pm(\vk+\vn\pi)\bigr)$,
$E_\pm=E_\pm(\vk)$ given in Eq.\ (\ref{Epm}).  With Eqs.\ (\ref{A2}) and
(\ref{reg1}) we therefore obtain
\eqa
{\rm Tr}\log G^{-1} =
\nonu
\sum_{\vk\in \Lambda_{\pi/2}^* }
\sum_{\vn}\sum_{\sigma=+,-}\beta \Ln\bigl(E_\sigma (\vk+\vn\pi) \bigr) \nonu
= \sum_{\vk\in \Lambda_{\pi}^* }
\sum_{\sigma=+,-} \beta \Ln\bigl(E_\sigma (\vk)\bigr)
\eqaend
where
\begin{mathletters}
\eq
\Lambda_{\pi}^*=\bigl\{ \vk=(k_1,\ldots, k_d)| -\pi\leq k_i <\pi,\;
L k_i/2\pi\in \Z   \bigr\}
\eqend
is the Brillouin zone of the full lattice $\Lambda_L$.
Thus $f_0^L=\cF/\beta L^d$, $\cF$ Eq.\ (\ref{F}), is
\eq
\label{fL}
f_0^L = \frac{r^2+s^2}{U} - \frac{1}{L^d}
\sum_{\vk \in \Lambda_{\pi}^* } [ \Ln(E_+ ) + \Ln( E_- ) ]\: .
\eqend
\end{mathletters}Since $\frac{1}{L^d} \sum_{\vk \in \Lambda_{\pi}^* }$
in the limit
$L\to\infty$ becomes $\intdk$, we obtain $f_0^\infty=f_0$ Eq.\ (\ref{f0}).
This concludes our proof.

\section{Saddle point evaluations}

In this appendix we justify in detail the saddle point evaluation of our
integral Eq.\ (\ref{eq}). We first prove Eq.\ (\ref{f}), and then evaluate
$\mu(V)$ for $\xx_1<\xx<\xx_2$.

\subsection{The $r$-integral}\label{B1}
For finite $L$, the HS action Eq.\ (\ref{F}) for the N\'eel configurations
Eq.\ (\ref{Neel}) equals $\cF=\beta L^d f_0^L$ with $f_0^L$ given in Eq.\
(\ref{fL}).  We consider
\eq
{\cal I}^L ({\cal C},V ) = \int_{{\cal C}} {d} r\,
e^{-V  f_0^L(r,s) }
\eqend
with ${\cal C}$ some integration path in the complex $r$-plane.  We are
interested in this integral for $V \to\infty$ and the integration path
along the real $r$-line,
\eq
{\cal C}_{real}\, : \, r_{real}(\tau) = \tau ,
\quad -\infty\leq \tau\leq \infty \: .
\eqend
For the saddle point evaluation of this integral below we need to show that
one can deform ${\cal C}_{real}$ to a path of steepest descent ${\cal
C}_{st.d}$ through the dominating saddle point of $f_0$.  This is
non-trivial since $\Ln(z)$ has branch points in $z={i} (2n+1)\pi\beta$, $n$
integer.  However, $e^{\beta \Ln(z) } = 2\cosh(\beta z/2)$, thus we see from
Eq.\ (\ref{fL}) that $e^{-\beta L^d f_0^L(r,s) }$ is analytic for all $r\neq
\infty$, and ${\cal I}^L({\cal C}_{real} ,\beta L^d) = {\cal I}^L({\cal
C}_{st.d.}, \beta L^d)$ follows from Cauchy's theorem.  Moreover, since
$f_0^L = f_0^\infty + {\cal O}(1/L)$, we get
${\cal I}^\infty ({\cal C}, V ) = {\cal I}^L ({\cal C}, V )\,
e^{{\cal O}(V/ L ) } $. We thus conclude that
\eq
\label{aha}
{\cal I}^\infty ({\cal C}_{real} ,\beta L^d ) =
{\cal I}^\infty ({\cal C}_{st.d.}, \beta L^d )\,
e^{{\cal O} (\beta L^{d-1} )} \: .
\eqend

In the following we consider ${\cal I}^\infty({\cal C},V)$.  Introducing
the DOS $N(E) = \intdk [ \delta(E-w) + \delta(E+w) ]$ with
$w=\sqrt{s^2 + \epsilon(\vk)^2 }$ we write $f_0^\infty=f_0$ as
\eq
f_0(r) = \frac{s^2+r^2}{U} - \int {d} E N(E) \Ln (E+ ir - \mu) \, .
\eqend
The saddle point equation $\partial f_0/\partial r=0$ has a single purely
imaginary solution $r^*={i} y^*$, $y^*$ real, where
\eq
y^* = -\frac{U}{2}\int {d} E N(E) f_\beta (E-y^*-\mu) \, .
\eqend
To determine the path ${\cal C}_{st.d}$ we evaluate
\begin{mathletters}
\eq
f_0(r^*+\xi) = f_0(r^*) + a\xi^2 - {i} b\xi^3 +{\cal O}(\xi^4)
\eqend
where
\eqa
a=\frac{1}{U} +  \int {d} E N(E)\frac{\beta}{8\cosh^2\bigl(\beta
(E-y^*-\mu)/ 2\bigr)} \nonu
b = \int {d} E N(E)\frac{\beta^2 \tanh\bigl(\beta (E-y^*-\mu)/2
\bigr)}{12\cosh^2\bigl(\beta (E-y^*-\mu)/2 \bigr) } \: .
\eqaend
\end{mathletters}
>From this we can determine the path of steepest descent $r^*(\tau)= \tau +
{i} y^*(\tau)$ obeying $\Im f_0\bigl( r^*(\tau) \bigr)=0$,
\eq
y^*(\tau) = y^* + \frac{b}{2a} \tau^2 + {\cal O}(\tau^4) \: .
\eqend
Note that $y^*(\tau) = y^*(-\tau)$. We now choose
\eq
{\cal C}_{st.d}:\, r_{st.d.}(\tau) = \left\{ \bma{cc} \tau + {i} y^*(\tau)&
\mbox{ for $|\tau|\leq \varepsilon$ } \\ \tau + {i} y^*(\varepsilon)&
\mbox{ for $|\tau|\geq \varepsilon$ } \ema \right.
\eqend
for some $\varepsilon>0$. We obviously have [$r^*=r^*(0)$]
\[
f_0\bigl( r^*(\tau) \bigr) =  f_0(r^*) + a\tau^2 + {\cal O}(\tau^4)
\quad \mbox{ for $|\tau|\leq \varepsilon$ } \: .
\]
Moreover, by a simple calculation
\nonueqa
\Re\bigl( f_0(x+{i} y) - f_0({i} y) \bigr) =
\frac{x^2}{U} - \int{d} E N(E)
\nonu \times
\frac{1}{2\beta}\log\Bigl(1-\frac{\sinh^2(\beta x/2)}{\cosh^2(\beta (E-y-\mu)/
2) } \Bigr)
\geq \frac{x^2}{U}  \, ,
\nonueqaend
thus $a>1/U$ gives the estimate (for $\varepsilon$ sufficiently small)
\eq
\Re\Bigl( f_0\bigl(r_{st.d.}(\tau) \bigr) - f_0(r^*) \Bigr) \geq \frac{U}{2}
\tau^2\quad \forall \tau \: .
\eqend
Thus ${\cal I}^\infty({\cal C},V) = e^{-V f_0(r^*)}(\cdots) $ with
$|(\cdots)|\leq \int_{-\infty}^\infty d\tau \, e^{-V U \tau^2/2}$.
Combining this with Eq.\ (\ref{aha}) gives
\eq
-\lim_{V\to\infty} \frac{1}{V}
\log\bigl( {\cal I}^\infty ({\cal C}_{real},\beta L^d) \bigr)
= f_0(r^*)
\eqend
equivalent to Eq.\ (\ref{f}).

\subsection{Degenerate saddle points}\label{B2}
For large but finite $V$ and $\mu$ close to $\mu_c$ the integral Eq.\
(\ref{eq}) equals
\eq
Z \simeq e^{-V \Omega_1} + e^{-V\Omega_2}
\eqend
up to correction terms which are irrelevant for $V\to\infty$ and thus will
be ignored.  The first and second terms here are the contributions from the
first and second saddle points $s_1^*=0$ and $s_2^*=s^*$, respectively,
\eq
\label{Omi}
\Omega_i(\mu) = f_0^*(s^*_i,\mu) + c_V(s^*_i,\mu) \: .
\eqend
The contributions
\[
c_V(s^*_i,\mu) \simeq \frac{1}{V}\log\bigl(A_i(\mu) V \bigr)
\qquad (A_i>0)
\]
come from Gaussian integrations in the regions close to the saddle points.
The precise form of the $c_V$ actually is not important for us and we will
only use that they vanish for $V\to\infty$.  Using $Z=e^{-V\Omega}$ and
$\xx=-\partial\Omega/\partial \mu$ we get
\eq
\xx \simeq \frac{ e^{-V \Omega_1}\, \xx_1  + e^{-V\Omega_2}\, \xx_2}
{ e^{-V \Omega_1} + e^{-V\Omega_2} }
\eqend
where $\xx_i=-\partial\Omega_i/\partial \mu$.  This is precisely of the
form Eq.\ (\ref{weight}) if we choose $ w/(1-w) = (\xx_2 -\xx)/(\xx-\xx_1)
\simeq \exp{\bigl(-V (\Omega_1-\Omega_2)\bigr) }$. For
$\mu=\mu_c+\delta\mu$ we have $\Omega_1-\Omega_2 \simeq
-(\xx_1-\xx_2)\delta\mu + [c_V(s_1^*,\mu_c)-c_V(s_2^*,\mu_c)]$ by
definition of $\mu_c$, thus we get
\eq
V\delta\mu \simeq
-\frac{1}{(\xx_2-\xx_1)}\log\Bigl( \frac{\xx_2 -\xx}{\xx-\xx_1} \Bigr)
\eqend
up to terms which vanish for $V\to\infty$.  This shows that
$\delta\mu={\cal O}(1/V)$ can be adjusted such as to produce arbitrary
doping $\xx$ between $\xx_1$ and $\xx_2$, as claimed in the text.
\end{appendix}


\begin{references}

\bibitem{Hubbard} M.C.\ Gutzwiller, Phys.\ Rev.\ Let..\ {\bf 10}, 159 (1963);
J.\ Hubbard, Proc.  Roy.  Soc.  London A {\bf 276}, 238 (1963); K.\
Kanamori, Prog.  Theor.  Phys.  {\bf 30}, 275 (1963)

\bibitem{Anderson} P.W.\ Anderson, Science {\bf 235}, 1196 (1987)

\bibitem{Review} E.\ Dagotto, Rev.\ Mod.\ Phys.\ {\bf 66}, 763 (1994)

\bibitem{Anderson0} P.W.\ Anderson, {\em Concepts in Solids},
Frontiers in Physics, Addison-Wesley (1963)

\bibitem{BL} V.\ Bach, E.H.\ Lieb and J.P.\ Solovej, J.\ Stat.\ Phys. {\bf 76},
3 (1994)

\bibitem{walls} J.\ Zaanen and O.\ Gunnarsson, Phys.\ Rev.\ B {\bf
40}, 7391 (1989)

\bibitem{HartreeFock} A.\ Singh and Z.\ Te$\check{\mbox{s}}$anovi\'c,
Phys.\ Rev.\ B {\bf 41}, 614 (1990); H.J.\ Schultz, Phys.\ Rev.\ Lett.\
{\bf 64}, 1445 (1990); D.\ Poilblanc and T.M.\ Rice, Phys.\ Rev.\ B {\bf 39},
9749 (1989); W.P.\ Su, Phys.\ Rev.\ B {\bf 37}, 9904 (1988)

\bibitem{spiral} E.\ Arrigoni and G.C.\ Strinati, Phys.\ Rev.\ {\bf B 44},
7455 (1991); H.J.\ Schultz in Ref.\ \cite{HartreeFock}

\bibitem{phasesep} V.J.\ Emery, S.A.\ Kivelson, H.Q.\ Lin,
Phys.\ Rev.\ Lett.\ {\bf 64}, 475 (1990); L.B.\ Ioffe and A.I.\ Larkin,
Phys.\ Rev.\ B {\bf 37}, 5730 (1988); P.B.\ Visscher, Phys.\ Rev.\ B {\bf
10}, 943 (1974); for review see Ref.\ \cite{Review}

\bibitem{VLLGB} J.A.\ Verg\'es, E.\ Louis, P.S.\ Lomdahl, F.\
Guinea and A.R.\ Bishop, Phys.\ Rev.\ B {\bf 43}, 6099 (1991)

\bibitem{Schulz} H.J.\ Schulz, Phys.Rev.Lett. {\bf 65}, 2462 (1990)

\bibitem{Wolff} see e.g.\ U.\ Wolff, Nucl.\ Phys.\ {\bf B225} [FS9], 391 (1983)

\bibitem{EK} V.J.\ Emery and S.A.\ Kivelson, Physica C 209, 597 (1993)


\bibitem{NO} See e.g.\ J.W.\ Negele and H.\ Orland,
{\em Quantum many-particle systems}, Frontiers in Physics,
Addison-Wesley (1988)

\bibitem{temp} The temperatures in Figs.~1 and 2 are given in units of $t$.
They were chosen such that for $t=0.5\,$eV we get $T=50$, $1000$ and 
$4000\,$Kelvin in Fig.~1, 2(a) and 2(b), respectively.

\bibitem{manfred} We thank M.\ Salmhofer for discussions on this point.

\end{references}
\end{document}